\documentclass[amssymb,amsmath,floatfix,twocolumn,superscriptaddress,prl]{revtex4}
\usepackage{graphicx}
\usepackage{lastpage}
\usepackage{amssymb}
\newcommand{\ket}[1]{|#1 \rangle}

\begin{document}

\title{High Performance Quantum Computing}
    \author{Simon J. Devitt}
    \email{devitt@nii.ac.jp}
    \address{National Institute for Informatics, 2-1-2 Hitotsubashi, Chiyoda-ku, Tokyo 101-8430, Japan.}
 %    \author{Austin G. Fowler}
 %   \address{Institute for Quantum Computing, University of Waterloo, Waterloo, Canada.}
% \author{Ashley M. Stephens}
 %   \address{Centre for Quantum Computing Technology, Department of Physics,
 %   University of Melbourne, Victoria, Australia}
 %     \author{Andrew D. Greentree}
 %   \address{Centre for Quantum Computing Technology, Department of Physics,
 %   University of Melbourne, Victoria, Australia}
 %       \author{Lloyd C.L. Hollenberg}
 %   \address{Centre for Quantum Computing Technology, Department of Physics,
 %   University of Melbourne, Victoria, Australia}
      \author{William J. Munro}
    \address{Hewlett-Packard Laboratories, Filton Road, Stoke Gifford, Bristol BS34 8QZ, United Kingdom}
      \address{National Institute for Informatics, 2-1-2 Hitotsubashi, Chiyoda-ku, Tokyo 101-8430, Japan.}
 %    \author{Jeremy L. O'Brien}
 %   \address{Centre for Quantum photonics, H. H. Wills Physics Laboratory \& Department of Electrical and Electronic Engineering, University of Bristol, Merchant Venturers Building, Woodland Road, Bristol, BS8 1UB, UK}
        \author{Kae Nemoto}
    \address{National Institute for Informatics, 2-1-2 Hitotsubashi, Chiyoda-ku, Tokyo 101-8430, Japan.}
\date{\today}

\begin{abstract}

The architecture scalability afforded by recent proposals of a large scale photonic based quantum computer, 
utilizing the theoretical developments of topological cluster states and the photonic chip, allows us to move 
on to a discussion of massively scaled Quantum Information Processing (QIP).  In this letter we introduce the 
model for a secure and unsecured topological cluster mainframe.  We consider the quantum analogue of 
High Performance Computing, where a dedicated server farm is utilized by many users to run 
algorithms and share quantum data.  The scaling structure of photonics based topological cluster computing leads 
to an attractive future for server based QIP, where dedicated mainframes can be constructed and/or 
expanded to serve an increasingly hungry user base with the ideal resource for individual quantum information 
processing.

\end{abstract}
 
\maketitle

Since the introduction of quantum information science in the late 1970's and early 1980's, 
a large scale physical device capable of high fidelity quantum information processing (QIP) 
has been a major and highly sought after goal.  While quantum information has lead to many extraordinary developments 
in foundational quantum theory, quantum atom/optics, solid state physics and optics many researchers world wide are 
still striving towards a physical quantum computer.  

The issue of computational scalability for QIP has been an intensive area of research for not only physicists but also 
computer scientists, mathematicians and network analysis and in the past decade has seen many proposals 
for scalable quantum devices for a variety of quantum architectures~\cite{arch}.  The complexity in designing a large scale 
quantum computer is immense and research in this area must incorporate complex ideas in theoretical and 
experimental physics, information theory, quantum error correction, quantum algorithms
and network design.  Due to the relative infancy of theoretical and experimental QIP it has been difficult to implement 
theoretically scalable ideas in quantum information theory, error correction and algorithm design into 
an architectural model where the transition from 1-100 qubits to 1-100 million qubits is conceptually 
straightforward.

Recent theoretical advancements in computational models for QIP has introduced an extremely elegant pathway to 
realize a enormously large QIP system in optics.  Topological cluster state computing, first introduced 
by Raussendorf, Harrington and Goyal~\cite{Raussendorf4} has emerged as an extremely promising computational model for QIP and 
integration of this model with chip-based photon/photon gates such as the photonic module has lead to 
a promising optical realization of 
quantum computation~\cite{Devitt1,Devitt2}.  
The conceptual scalability of the chip based topological optical computer allows, for the first time, 
a grounded discussion on large scale quantum information processing, beyond the individual computer.  
In this letter we take the scalability issue one step further, examining the possible long term 
implementation of topological cluster state computing with the photonic chip and discuss what the future may 
hold for this architectural model of QIP.

Traditional discussions of scalability in QIP is generally limited to the issue of constructing a single, moderately large scale 
quantum computer, capable of performing non-trivial algorithms for a single user.  In the case of topological 
cluster state computation in optics we can consider the possibility of 
client/mainframe quantum devices and start to consider the quantum analogue of classical high performance computing, 
namely High Performance Quantum Computing (HPQC) where a large, generic quantum resource is made available to multiple 
clients to perform independent (or joint) QIP.  

Topological cluster state computing in optics is uniquely suited to this task for several reasons.  Aside from the error correcting and 
resource benefits of the topological cluster model, the basic geometric structure of the lattice allows for multi-user computation that 
would be problematic when utilizing the more traditional 2D cluster state techniques~\cite{Raussendorf1}.  In traditional 2D cluster state 
computing, multiple users could not interact data with each other or with a central resource core without transporting 
quantum information through the cluster resource of other users.  Essentially multi-user interactions would be 
a Linear Nearest Neighbor (LNN) network.  Moving to 3D topological clusters convert this LNN network 
topology into a 2D grid, enabling the partitioning of the cluster lattice into user regions and 
resource regions.  
Additionally, as the lattice is carried by single photons we can potentially 
integrate a mainframe model with developments in quantum communications and entanglement distribution~\cite{ent}. 
This gives a layer of security to the HPQC which would be difficult, if not impossible to achieve for 
multi-user, matter based qubit architectures.   

\begin{figure*}[ht]
\begin{center}
\resizebox{130mm}{!}{\includegraphics{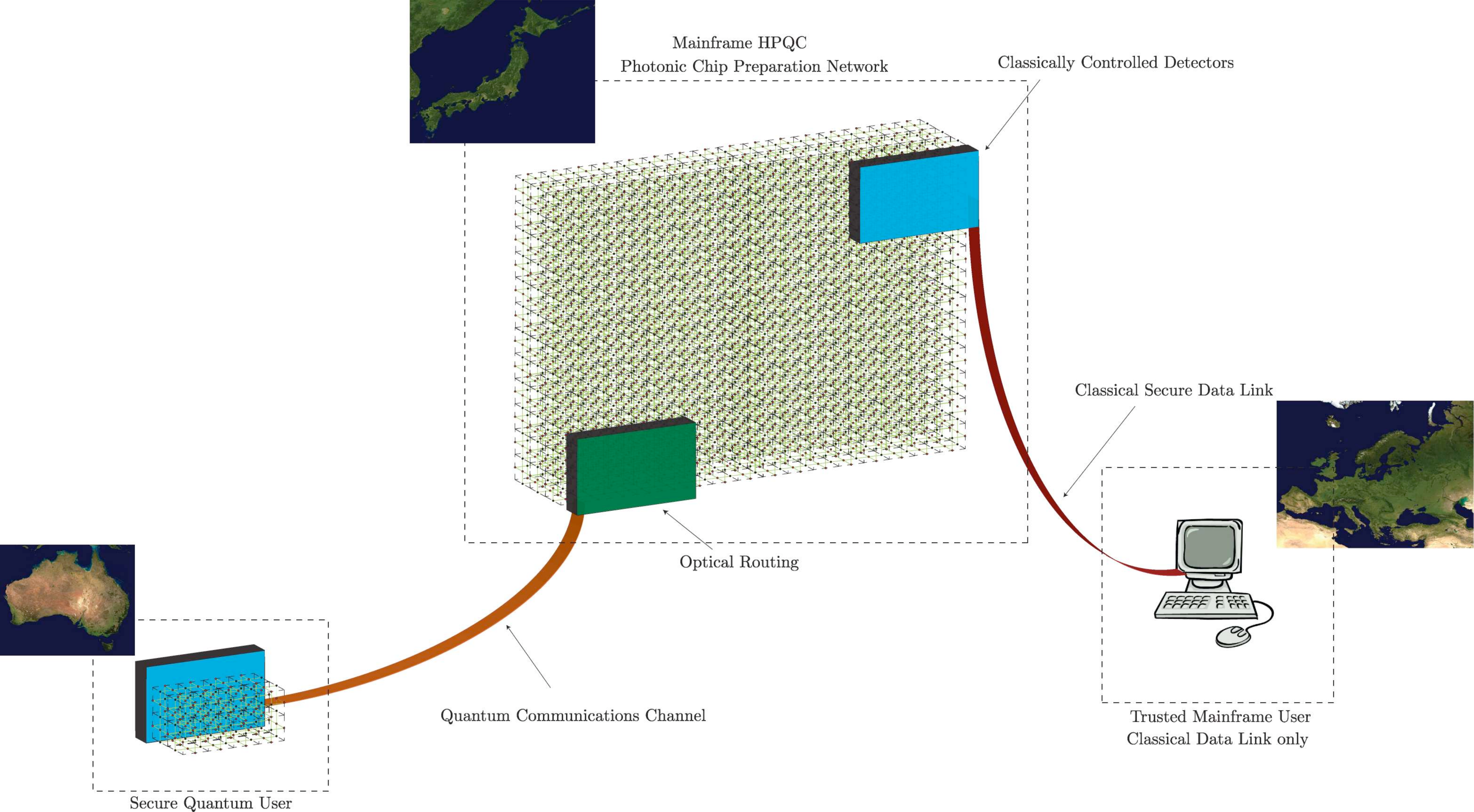}}
\end{center}
\vspace*{-10pt}
\caption{A central mainframe HPQC would 
consist of a massive cluster preparation network built from single photons sources and photonic chips.  Once the cluster is 
prepared, users can log on and perform individual computations in one of two ways.  A trusted mainframe model is where the 
user submits a classical data stream corresponding to the measurement pattern for a quantum algorithm.  The secured quantum user 
has access to a high fidelity quantum communications link between themselves and the mainframe.  The alloted portion of the global 
lattice is then physically routed to the user and photon measurements are performed locally.}
\label{figure:mainframe}
\end{figure*}

Here we introduce the basic framework for a potential HPQC based on topological cluster state computing in the photonic 
regime [Fig.~\ref{figure:mainframe}].  
We discuss two possible mainframe models, one where multi-user computation is performed locally by the 
mainframe and another where partitions of the mainframe lattice are sent via quantum communications channels to 
individual users.  We complete the discussion by providing a example of a partition structure for the mainframe lattice which 
satisfies many of the necessary components of a HPQC mainframe and give a basic estimate of the number of photonic chips 
required for a massive mainframe quantum server. 

The first model we consider we denote the trusted mainframe model.  This is where individual 
users connect via classically secure data pathways and the mainframe host is trustworthy.  Each client  
logs onto the host and transmits a classical data stream, corresponding to the desired quantum algorithm, to the 
host (via a sequence of photon measurement bases).  
The mainframe will then run the quantum algorithm locally and 
once the computation is complete, transmits the resulting classical information back to the user. 

This model has very substantial benefits.  First, each user does not require quantum communications 
channels or any quantum infrastructure locally.  All that is required is that each user compile a quantum algorithm into 
an appropriate classical data stream which is sent to the mainframe.  
Additionally, the host does not need to transmit any data to the user during computation.  All internal 
corrections to the lattice which arise due to its preparation and error correction procedures are performed 
within the mainframe.  The only data which is transmitted to the user is the classical result from the 
quantum algorithm.  Finally, as each user independently logs on to the system to run a quantum algorithm, the 
mainframe can be configured to assign resources dynamically.  If one user requires a large number of 
logical qubits and if the mainframe load is low, then the host can adjust to allocate 
a larger partition of the overall lattice to one individual user.  

While the user/mainframe interaction of this mainframe model is identical to classical 
models for high performance computing, the fact that we are working with qubits suggests the 
possibility of secure HPQC.  In the trusted mainframe model 
the classical data stream from the user to host is 
susceptible to interception (although quantum key distribution and secure data links can be utilized to 
mitigate this issue) and the quantum mainframe has complete access to both the quantum algorithm being 
run on the server and the results of the computation.  
If sensitive computation is required we can combine the mainframe with high fidelity communication 
channels to perform a secure version of HPQC in a manner unavailable to classical distributed computing.

As the topological lattice prepared by the mainframe is photon based, we are able to 
utilize high fidelity optical communications channels to physically transmit a portion of the 3D lattice to 
the client. 
Compared with the trusted mainframe model, this scheme has some technological disadvantages.  
High fidelity quantum communication channels are required to faithfully transmit entangled photons 
from the mainframe to each client.  While purification protocols could, in principal, be utilized to increase transmission fidelity, 
this would be cumbersome, and given that topological models for QIP exhibit very high thresholds (of the 
order of 0.1-1\%) it is fair to assume that communication channels will be of sufficient reliability when a 
mainframe device is finally constructed.
Secondly, each client must have access to a certain amount of quantum technology.  Specifically, 
a set of classically controlled, high fidelity single photons wave-plates and 
detectors.  This allows each client to perform their own measurement of the photon stream to perform computation locally.  

Security arises as the
quantum data stream never carries information related to the quantum algorithm 
being run on the client side.  As the photon 
stream transmitted to the client is the 3D topological lattice generated by the mainframe, interrogation of the 
quantum channel is unnecessary as the state transmitted is globally known.  
Additionally, the only classical information sent between mainframe and user is related to the initial 
eigenvalues of the prepared lattice (obtained from the mainframe preparation network),  no other classical 
data is ever transmitted to or from the user.  This implies that even if an eavesdropper successfully 
taps into the quantum channel and entangles their own qubits to the cluster they will not know 
the basis the user chooses to measure in or have access to the classical 
error correction record.   While an eavesdropper could employ a denial of service attack, the ability to 
extract useful information from the quantum channel is not possible without access to the classical information 
record measured by the client.

A second benefit to the secure model is that the client has ultimate control of whether their portion of the lattice 
generated by the host remains entangled with the larger global lattice of the mainframe.  Performing 
$\sigma_z$ basis measurements on any photon within the cluster simply disentangles it from the lattice.  
Hence if the mainframe transmits a partial section of the generated lattice to the client, they simply perform $\sigma_z$ 
basis measurements on all photons around the edge of their partitioned allotment and they are guaranteed that 
neither the host and/or other users sharing the mainframe lattice can interact their portion of the 
lattice with the clients alloted section.  This severing of the users sub-lattice from the mainframe would generally 
be recommended.  If the sub-lattice is still linked to the mainframe, error correction procedures would 
need to be co-ordinated with the mainframe and classical data continually exchanged.  This is due to the fact 
that error chains are able to bridge the region between user and host when links remain in-tact.

When a user has completed their task they have the option of making their results available to the global lattice, 
either to be utilized again or shared with other users.  If the client does not wish to share the final quantum state 
of their algorithm, they measure all defect qubits and restore their portion of the lattice to a defect free state.  
If however, they wish to make available a non-trivial quantum state to the mainframe, then after their quantum algorithm is 
completed they can cease to measure the photons on the boundary of their allotted lattice.  
Once the client logs off the system, the quantum state of the defect qubits within this lattice will remain (provided the 
mainframe automatically continues measuring the sub-lattice to enact identity operations).  
Consequently, at a later time, the original user may decide to log onto the system again, or a second user may choose to 
log on that sub-lattice and continue to manipulate the stored data as they see fit (note that it is assumed that the 
global lattice is of sufficient size to allow for significant error protection and hence long term information storage).  
Additionally, this same methodology can be utilized to allow different users to interact quantum states.  As with 
the previous case, two users may decide to perform independent, private, quantum algorithms up to some finite time and then 
interact data.  Each user then ceases severing the connections to the global lattice and receives half an encoded 
Bell state from the mainframe, allowing for the implementation of teleportation protocols.  

Although the preparation of a large 3D cluster lattice with photonic chips has been examined, how to 
partition resources for an optimal, multi-user device is a complicated networking problem.  At this 
stage we will simply present an example partition structure for the resource lattice, hopefully demonstrating 
some of the essential features that would be needed for this model.  We will approach this analysis with 
some basic numerical estimates to give an idea of the resource costs and physical lattice sizes for 
a mainframe device. 
\begin{figure}[ht]
\begin{center}
\resizebox{75mm}{!}{\includegraphics{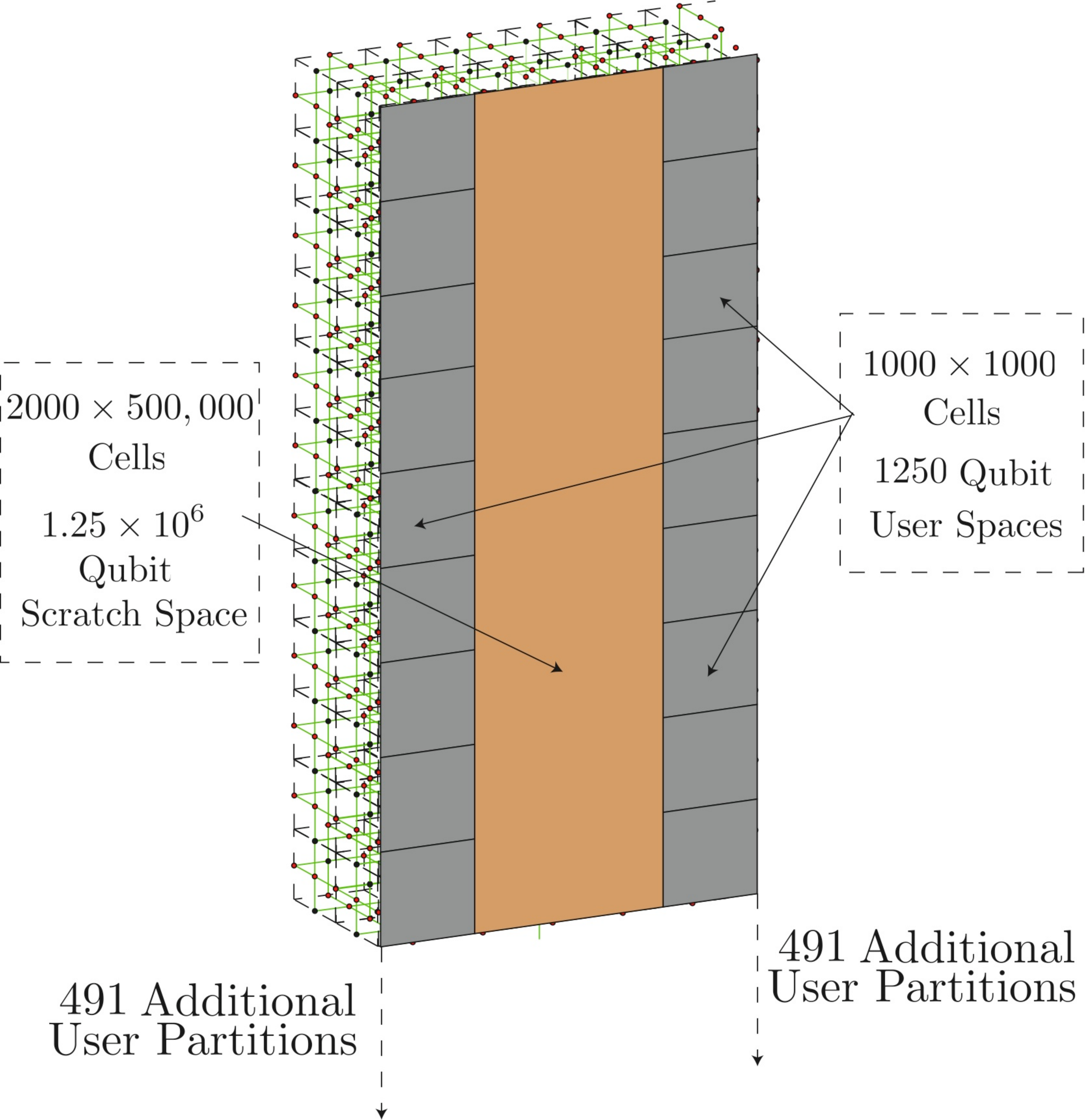}}
\end{center}
\vspace*{-10pt}
\caption{Illustrated is an 
example partitioning of the global 3D lattice for a HPQC mainframe.  This global lattice measures 
$4000\times 500,000$ unit cells and requires approximately $7.5\times 10^9$ photonic chips to prepare.  If utilized as a single 
cluster computer, 2.5 million logical qubits are available with sufficient topological protection for approximately $10^{16}$ logical 
operations (where a logical operation is defined as the measurement of a single unit cell).}
\label{figure:partition}
\end{figure}

The HPQC mainframe will consist of two regions, an outer region corresponding to user partitions and an inner region which we will 
denote as scratch space.  The scratch space will be utilized to for two primary tasks.  
The first is to provide logical Bell states to individual users in order to interact quantum information, the second is 
to distill and provide the high fidelity logical ancillae states $\ket{A} = (\ket{0} + i\ket{1})/\sqrt{2}$ and 
$\ket{Y} = (\ket{0} + \exp(i\pi/4)\ket{1})\sqrt{2}$ which are needed to enact non-trivial single qubit rotations 
that cannot be directly implemented in the topological model.  Purifying these 
states is resource intensive and as these states are required often for a general quantum algorithm it would be 
preferable to have an offline source of these states which does not consume space on the user partitions.  

It should be stressed that the size of the scratch space lattice will be heavily dependent on the fundamental 
injection fidelity of these non-trivial ancilla states and consequently the amount of required state distillation.  
This illustrative partitioning of the mainframe lattice, shown in Fig.~\ref{figure:partition} 
allocates a scratch space of $1000\times 1000$ cells for each user region (effectively another computer the of the same size).
In general, state distillation of ancilla states requires a large number of low fidelity qubits and distillation cycles and 
users will require a purified ancilla at each step of their computation~\cite{Fowler}.  Therefore, the scratch space 
could be significantly larger than each user partition.  
This does not change the general structure of the lattice partitioning, instead the width of the central scratch region is 
enlarged with user partitions still located on the boundaries.  The primary benefit of requiring the mainframe to 
prepare purified ancillae is dynamical resource allocation, performed at the software level by the 
mainframe.  By allowing the mainframe to prepare all distilled 
ancillae it is able to adjust the user/scratch partition structure to account for the total number of 
users and the required preparation rate of distilled states. 

Based on this partitioning of the mainframe lattice we can illustrate the resource costs through a basic 
numerical estimate.  As shown in~\cite{Devitt2}, under reasonable physical assumptions, a large scale 
topological computer capable of approximately $10^{16}$ logical operations (where a logical operation is 
defined as the measurement of a single cluster cell) requires 
approximately 3000 photonic chips per logical qubit, measuring $20\times 40$ cells in the lattice.  
We therefore allocate each user a square region 
of the overall lattice measuring 1000$\times 1000$ unit cells, containing $50\times 25$ logical qubits and requiring 
approximately $3.75\times 10^6$ photonic chips to prepare.  
Additionally we consider a HPQC mainframe of sufficient size to accommodate 1000 
individual user regions of this size with a scratch space 
two user regions wide and 500 user regions deep. Hence, this HPQC will need to generate a rectangular lattice measuring 
$4000\times 500,000$ cells and require of order $7.5\times 10^9$ photonic chips to prepare.  

This may seem like a extraordinary number of devices to manufacture and incorporate into a large scale lattice generator, 
but one should recognize the enormous size of this mainframe.  The partition structure is determined at the software level, no 
changes to the lattice preparation network is required to alter the structure of how the lattice is utilized.
Hence, if desired, this mainframe can be utilized as a single, extremely large, quantum computer, 
containing 2.5 million logical qubits, with topological protection for approximately $10^{16}$ operations, 
more than sufficient to perform {\em any} large scale quantum algorithm or simulation ever proposed. 

In conclusion, we have introduced the concept of the High Performance Quantum Computer, where a massive 
3-dimensional cluster lattice is utilized as a generic resource for multiple-user quantum information processing.  
The architectural model of 3D topological clusters in optics allows 
for the conceptual scaling of a large topological cluster mainframe well beyond what could theoretically 
be done with other architectures for QIP. 
As an example we illustrated a possible lattice partitioning of the mainframe system.  This partition, while not optimal, 
shows some of the necessary structures that would be required for multi-user 
quantum computing.  With this partition structure we were able to estimate the number of 
photonic chips required to construct a mainframe device.  The construction of approximately 7.5 billion 
photonic chips leads to an extraordinary large multi-user quantum computer.  
While this is certainly a daunting task, this sized computer would represent the ultimate goal 
of QIP research that began in the late 1970's.  

The authors thank A.G. Fowler, N. L\"{u}tkenhaus, A. Broadbent and T. Moroder for helpful discussions.
The authors acknowledge the support of MEXT, JST, HP and the EU project QAP.


\begin{thebibliography}{00}
\bibitem{arch}
D.~Kielpinski, C.~Monroe, and D.~J.~Wineland.
\newblock {\em Nature} \textbf{417}, 709 (2002);
J.~M.~Taylor \emph{et. al.},
\newblock {\em Nature Phys.} \textbf{1}, 177 (2005);
L.~C.~L.~Hollenberg, A.~D.~Greentree, A.~G.~Fowler, and C.~J.~Wellard.
\newblock {\em Phys. Rev. B} \textbf{74}, 045311 (2006);
A.~G. Fowler \emph{et. al.}
\newblock {\em Phys. Rev. B} \textbf{76}, 174507 (2007);
R. Stock and D.F.V. James,
\newblock {\em arXiv:0808.1591} (2008).

\bibitem{Raussendorf4}
R. Raussendorf and J. Harrington,
\newblock {\em Phys. Rev. Lett.} \textbf{98}, 190504 (2007);
R. Raussendorf, J. Harrington and K. Goyal,
\newblock {\em New J. Phys.} \textbf{9}, 199 (2007);
A.G. Fowler, A.M. Stephens and P. Groszkowski,
\newblock {\em arXiv:0803.0272} (2008);
A.G. Fowler and K. Goyal,
\newblock {\em arXiv:0805.3202} (2008).

\bibitem{Devitt1}
S.J. Devitt \emph{et. al.},
\newblock {\em Phys. Rev. A.} \textbf{76}, 052312 (2007);

\bibitem{Devitt2}
S.J. Devitt \emph{et. al.},
\newblock {\em arXiv:0808.1782} (2008).

\bibitem{Raussendorf1}
R. Raussendorf and H.J. Briegel,
\newblock {\em Phys. Rev. Lett.} {\bf 86}, 5188 (2001).

\bibitem{ent}
P. Villoresi \emph{et. al.}
\newblock {\em New. J. Phys.} {\bf 10}, 033038 (2008);
R. Ursin \emph{et. al.}
\newblock {\em Proc. 2008 Microgravity Sciences and Process Symposium} (2008);
SECOQC Project
\newblock www.sqcoqc.net.

\bibitem{Fowler}
A.G. Fowler,
\newblock {\em quant-ph/0506126} (2005).


\end{thebibliography}
\end{document}